\documentclass[a4paper,11pt]{article}
\usepackage{pos}

\input ./belle2sym.tex

\title{Results and Prospects of Radiative and Electroweak Penguin Decays at Belle II}

\author*[1]{Yo Sato}

\affiliation[]{Tohoku University,\\
  Miyagi, Japan}

\note{On behalf of the Belle II collaboration}

\emailAdd{yosato@epx.phys.tohoku.ac.jp}

\abstract{Radiative and electroweak penguin mediated decays of \B mesons are a great probe for physics beyond the Standard Model of particle physics. Furthermore, recently anomalies on exclusive $\b \to \s \ellell$ processes, which may imply lepton flavor universality violation, have been observed. Belle II experiment sheds light on the anomalies with the measurement of an inclusive analysis in this mode. Thanks to smaller hadronic uncertainties compared with exclusive modes, complementary information can be provided. Belle II is also a unique experiment to search for processes involving neutrinos such as $\B \to K^{(*)} \nunub$. We report the results and prospects of an inclusive $\B \to X_s \ellell$ analysis and a search for $\B \to K^{(*)} \nunub$.}

\FullConference{%
  40th International Conference on High Energy physics - ICHEP2020\\
  July 28 - August 6, 2020\\
  Prague, Czech Republic (virtual meeting)
}

\begin{document}
\maketitle

\section{Introduction}
In the Standard Model (SM) of particle physics, the flavor-changing neutral currents (FCNC) $b \to s$ and $b \to d$ proceed, mainly, via loop diagrams and thus are suppressed.
Since new heavy particles might be able to enter the loops, the branching fractions and angular observables for such processes are a great probe for physics beyond the SM.
Radiative and electroweak penguin \B meson decays are theoretically and experimentally clean FCNC processes compared with fully hadronic decays, because the final states include color singlet particle or particles, $\gamma, \ellell, \nunub$.
Furthermore, the anomalies in exclusive $\B \to K^{(*)}\ellell$ decays have been observed by the LHCb experiment \cite{Aaij:2015oid}  \cite{Aaij:2017vbb} \cite{Aaij:2019wad} \cite{Aaij:2020nrf} and more than 2$\sigma$ deviations in a similar direction were also reported by the Belle experiment \cite{Wehle:2016yoi} and the ATLAS experiment \cite{Aaboud:2018krd}.
Further studies on this field are therefore highly motivated.

The Belle II experiment contributes to this field by providing independent measurements to shed further light on the anomalies. The inclusive $\B \to X_s \ellell$ measurements at Belle II give complementary information with independent hadronic uncertainties.
Moreover Belle II is a unique experiment to be able to search for decays involving neutrinos, such as $\B \to K^{(*)} \nunub$. These processes are supposed to be discovered at early stage of data taking at Belle II.

In the following, we report the status and prospects of the radiative and electroweak penguin \B meson decays at the Belle II experiment.

\section{Measurements on the $b \to s \ellell$ process}
The measured anomalies in the exclusive $\B \to K^{(*)}\ellell$ decays, mentioned in the previous section, imply lepton flavor universality violation. 
Especially, the angular analysis of $\B \to K^* \ellell$ indicates smaller vector coupling constants in the muon modes, which is known as the Wilson coefficient $\mathcal{C}_{9}$. The difference between the SM expectation and the LHCb results is at the 3.3$\sigma$ level \cite{Aaij:2020nrf}.

The inclusive $\B \to X_s \ellell$ decay provides complementary information to the exclusive ones. The hadronic uncertainties of the inclusive decays are under better theoretical control with respect to those of the exclusive processes. 
Belle and BaBar have already performed measurements on $\B \to X_s \ellell$ \cite{Kaneko:2002mr} \cite{Iwasaki:2005sy} \cite{Sato:2014pjr} \cite{Aubert:2004it} \cite{Lees:2013nxa}. So far, all measurements are highly statistically limited due to the small cross section and limited data sample. 
Belle II is a unique experiment to achieve the measurements on $\B \to X_s \ellell$ with large statistics.

\begin{figure}[htbp]
  \centering
  \includegraphics[width=0.85\linewidth]{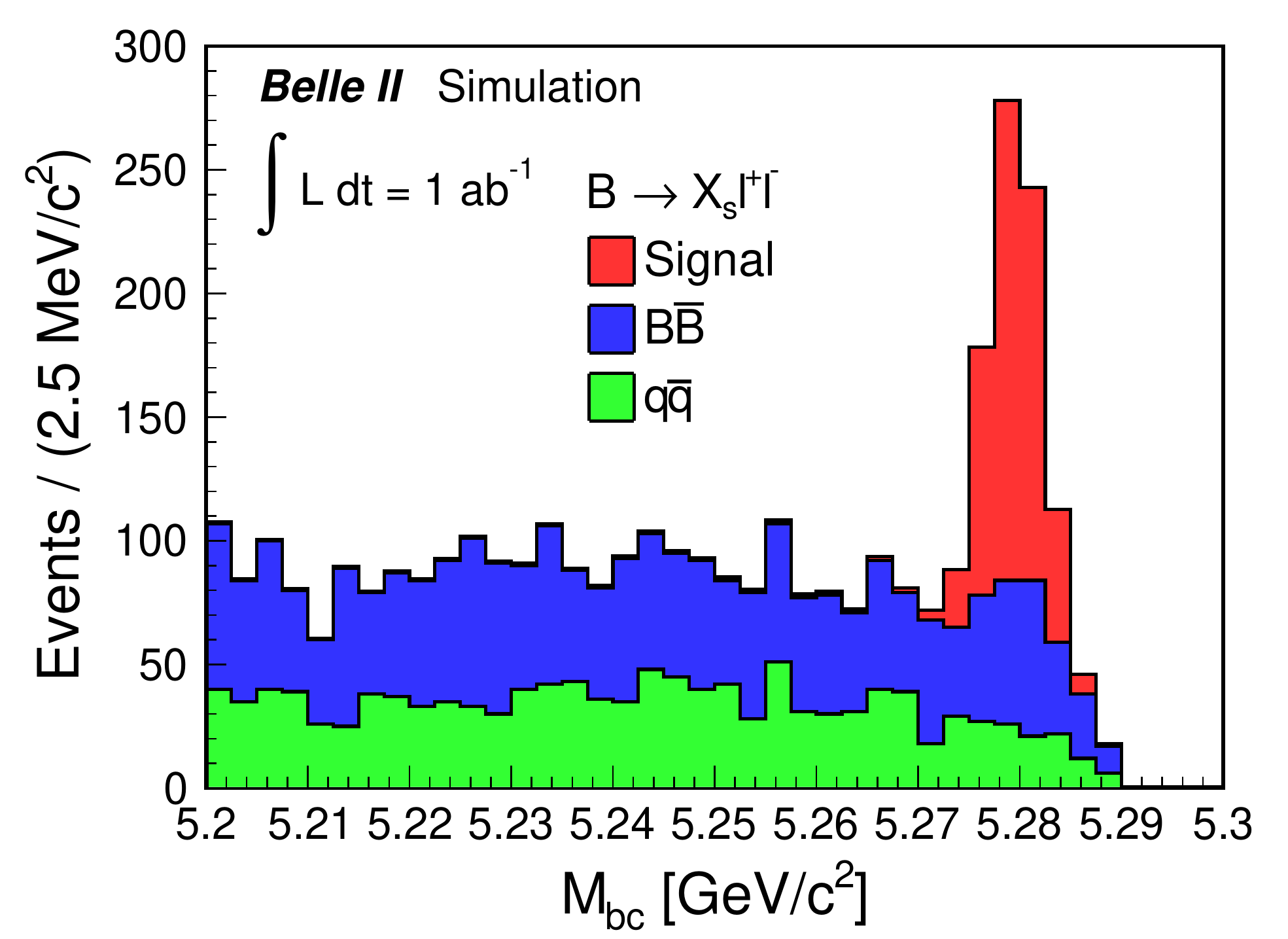}
\caption{ The beam-constrained mass ($M_{\rm bc}$) distribution of $\B \to X_s \ellell$ decay candidates with MC samples. The red histogram contains the signal process, the blue contains backgrounds from $\epem \to \BB$ process and the green contains backgrounds from $\epem \to \qqbar$ processes. }
\label{fig:MC_ll}
\end{figure}

The $\B \to X_s \ellell$ decays are reconstructed with a sum-of-exclusive approach. The hadronic system $X_s$ is reconstructed from $Kn\pi (0 \leq n \leq 4)$ and $3K$ final states allowing at most one $\pi^0$ and one $K_S^0$. The $X_s$ mass is required to be less than $2.0 \gevcc$ to suppress contamination from wrongly reconstructed $X_s$ candidates. Higher mass regions will be explored by using a larger dataset. 
One of the dominant background sources is $\B \to X_s \jpsi, X_s \psitwos$ followed by $\jpsi (\psitwos) \to \ellell$. These events are vetoed by applying cuts on the di-lepton invariant mass distribution. 
Other dominant backgrounds are \BB events with double semi-leptonic decay and \ccbar continuum events. They are suppressed by event shape variables, missing energy information, and vertex quality of \B meson. Another important background is $\B \to K m \pi (m\geq 2)$ with particle mis-identification of $\pipi$ as $\ellell$. 
Figure \ref{fig:MC_ll} shows the beam-constrained mass ($M_{\rm bc}$) distribution, $M_{\rm bc} = \sqrt{(E^*_{\rm beam})^2 - (p^*_B)^2}$, where $E^*_{\rm beam}$ is the beam-energy in the centre-of-mass (CM) system of $\epem$ and $p^*_{B}$ is the CM momentum of reconstructed \B meson. The signal peak on the $M_{\rm bc}$ distribution is expected to be significant over backgrounds thanks to the efficient background suppression \cite{Collaboration:2005}. The backgrounds which are able to make a peak on the $M_{\rm bc}$ distribution are estimated with a data-driven method. 
The Belle II analysis on $\B\to X_s \ellell$ is being finalized and first results will be shown in the near future.

The sensitivity on the branching fraction and the forward-backward asymmetry with an integrated luminosity of 50 \invab is estimated to be on a few percent level \cite{Kou:2018nap}.
The constraint on the Wilson coefficients $\mathcal{C}_9$ and $\mathcal{C}_{10}$ from these measurements will exclude the SM with 5$\sigma$ if the true values of them are at the current best fit \cite{Huber:2020vup}.

Also a fully-inclusive reconstruction method of the $\B \to X_s \ellell$ decay is currently being explored with dedicated simulation studies. The Full Event Interpretation (FEI) algorithm \cite{Keck:2018lcd} contributes to improve the tagging efficiency on the other side \B meson.

\section{Search for the $b \to s \nunub$ process}
The $b \to s \nunub$ processes are cleaner than their $b \to s \ellell$ counterparts because there are no photon mediated contribution on the leptonic part of the decay. 
In the SM, only the left-handed effective operator is relevant for the processes since \W bosons couple only to left handed fermions. 
New physics might be able to change the Wilson coefficient of the left-handed operator $\mathcal{C}_{L}$ and/or induce the right-handed contribution and thus make non-zero $\mathcal{C}_{R}$.

The reconstruction of $b \to s \nunub$ is experimentally extremely challenging due to neutrinos in the final state, which can not be detected. So far the process has never been observed. An upper limit on the branching fraction of $\B \to K^{(*)} \nunub$ was set by Belle and BaBar at the order of $10^{-5}$ \cite{Grygier:2017tzo} \cite{Lees:2013kla}, while the SM expectation is of the order of $10^{-6}$ \cite{Buras:2014fpa}. 

Belle II can observe the $\B \to K^{(*)} \nunub$ at early stage with only a few \invab of data, expecting the SM branching fractions. Belle II uses a hadronic and semi-leptonic tag with the FEI algorithm \cite{Keck:2018lcd} to reconstruct the other side \B meson.

According to simulation studies, a 10\% level sensitivity on the branching fraction, with an integrated luminosity of 50 \invab, is expected.
For the $\B \to K^{*} \nunub$ process, an angular analysis can be done to obtain the longitudinal polarization $F_L(K^*)$. The precision on the $F_L$ measurement is expected to be 20\%. 
Measurements of $\B \to K \nunub$ and $\B \to K^* \nunub$ can be combined to set constraints on the Wilson coefficient of $\mathcal{C}_L$ and $\mathcal{C}_R$. A large portion of the currently allowed region will be excluded.

First results on $\Bp \to K^+ \nunub$ using the current Belle II data are expected to be published soon. Despite of less statistics compared with Belle and BaBar, a new analysis method with inclusive tagging may yield in a competitive result.

\section{Summary}
Radiative and electroweak penguin \B meson decays are sensitive to physics beyond the SM. The recent anomalies highly motivate further studies on these channels. Belle II is a unique experiment to perform the measurements on the inclusive processes and to search modes involving neutrinos. 

The $\B \to X_s \ellell$ process is analyzed by reconstructing $X_s$ with the sum-of-exclusive method. Detailed MC simulation studies reveal the ability to suppress large amount of backgrounds and to make the signal peak on $M_{\rm bc}$ distribution significant. The expected sensitivity on the branching fraction and the forward-backward asymmetry measurements will exclude the SM with $5\sigma$ if the true values of $\mathcal{C}_9$ and $\mathcal{C}_{10}$ are at current best-fit. 

The $\B \to K^{(*)} \nunub$ process has not been observed yet. Assuming the branching fractions are at the SM expectation, Belle II can observe these processes even at an early stage. Combining semi-leptonic and hadronic tag methods, precise measurements on the branching fraction at the 10\% level and the longitudinal polarization $F_L(K^*)$ at the 20\% level will be possible with the expected full luminosity of Belle II. Measurements may exclude a large portion of the allowed parameter space of $\mathcal{C}_L$ and $\mathcal{C}_R$. 
Searches for $\Bp \to \Kp \nunub$ with an inclusive method are underway.

\section*{Acknowledgement}
Y.S. is supported by JSPS KAKENHI Grant Number JP19J10179 and Graduate Program on Physics for the Universe (GP-PU), Tohoku University.

\bibliography{belle2}
\bibliographystyle{belle2-note}



\end{document}